\documentclass[prl,aps,twocolumn,floatfix,showpacs]{revtex4}
\usepackage{graphicx,graphics,psfrag,amsmath,calc}
\usepackage{epsfig}
\begin{document}
\title{Disorder effects during the evolution from BCS to BEC superfluidity}
\author{Li Han and C. A. R. S{\'a} de Melo}
\affiliation{School of Physics, Georgia Institute of Technology, Atlanta, Georgia 30332, USA}
\date{\today}

\begin{abstract}
We describe the effects of disorder on the critical temperature of s-wave superfluids
from the BCS to the BEC regime, with direct application to ultracold
Fermi atoms.
In the BCS regime the pair breaking and phase coherence temperature scales
are essentially the
same allowing strong correlations between the amplitude and
phase of the order parameter.
As non-pair breaking disorder is introduced the largely overlapping
Cooper pairs conspire to maintain phase coherence such that the critical temperature
remains essentially
unchanged. However, in the BEC regime the pair breaking and phase coherence temperature
scales are very different such that non-pair breaking disorder can affect dramatically
phase coherence, and thus the critical temperature, without the requirement
of breaking tightly-bound fermion pairs simultaneously.
Finally, we find that the superfluid is more robust against weak
disorder in the intermediate region between the two regimes.
\end{abstract}
\maketitle

%
%% Introduction
%

Ultracold atoms are special systems for studying superfluid
phases of fermions or bosons at very low temperatures, because of unprecedented
tunability. In particular, ultracold fermions with
tunable interactions were used to study experimentally the so-called BCS-to-BEC
evolution, and population imbalanced systems. In addition, there are other interesting
directions to be pursued, including studies of
the BCS-to-BEC evolution in optical lattices~\cite{esslinger-06, ketterle-06},
and the effects of disorder during the BCS-to-BEC evolution, which would allow
the very important study of the simultaneous effects of interactions and
disorder at zero~\cite{orso-shklovskii-08} and
finite temperatures~\cite{sademelo-08}.

In ordinary condensed matter (CM) systems the control of interactions is not possible,
and the control of disorder is very limited, because the disorder potential is
not known and can not be changed at the turn of a knob. Thus, in standard CM
the disorder is usually described in terms of defects or impurities, whose
positions in the solid are assumed to be random.
In ultracold atoms it is now possible to create controlled disorder using laser
speckles or lasers with incommensurate wavelengths, which were
used to study the phenomenon of Anderson localization
in ultracold Bose atoms~\cite{aspect-08, modugno-08}, but that could also be used
to study disorder effects in ultracold fermions.

Thus, here, we describe the finite temperature phase diagram of
three dimensional (3D) s-wave Fermi superfluids from the BCS to the
BEC limit as a function of disorder, which is independent of the
hyperfine states of the atoms and is created by a
Gaussian-correlated laser speckle potential. Our main results are as
follows.  First, in the BCS limit the amplitude and phase of the
order parameter are strongly coupled, such that pair breaking and
loss of phase coherence occur simultaneously. In this case, the
critical temperature is essentially unaffected by weak disorder,
since the disorder potential is not pair-breaking and phase
coherence is not easily destroyed in accordance with Anderson's
theorem~\cite{anderson-59}. Second, in the BEC limit the breaking of
local pairs and the loss of phase coherence occur at very different
temperature scales. In this case, the critical temperature is
strongly affected by weak disorder, since phase coherence is more
easily destroyed without the need to break local pairs
simultaneously, and Anderson's theorem does not apply. Third, we
find that superfluidity is more robust to disorder in the
intermediate region between the BCS and BEC regimes.

%
%% Hamiltonian
%
To investigate the physics described above, we
start with the real space Hamiltonian ($\hbar = 1$) density for three
dimensional s-wave superfluids
\begin{equation}
\label{eqn:hamiltonian-real-space} {\cal H} ({\bf x}) =
\psi^{\dagger}_{\sigma} ({\bf x}) \left( -\frac{\nabla^2}{2m} - \mu
+ V_{dis} ({\bf x})\right) \psi_\sigma ({\bf x}) + {\hat U} ({\bf
x}),
\end{equation}
where
$ {\hat U} ({\bf x}) = + \int d{\bf x}^{\prime}  V ( {\bf x}, {\bf
x}^{\prime} ) \psi^{\dagger}_\uparrow ({\bf x}^{\prime})
\psi^{\dagger}_\downarrow ({\bf x}^{\prime}) \psi_{\downarrow}({\bf
x}) \psi_{\uparrow} ({\bf x}) $ %%
represents a term containing the interaction potential $V ({\bf x},
{\bf x}^{\prime}) = -g\, \delta ( {\bf x} - {\bf x}^{\prime} )$, and
$\psi^{\dagger}_{\sigma} ({\bf x})$ represents the creation of
fermions with mass $m$ and hyperfine state (spin) $\sigma$. In
addition, $V_{\rm dis} ({\bf x})$ represents the disorder potential,
and $\mu$, the chemical potential. We choose $V_{\rm dis} ({\bf x})$
to be independent of the hyperfine state, a choice that can be
easily relaxed.

Although there are many ways to introduce disorder, we will make a
particular choice that the disorder potential is governed by a
Gaussian distribution $ P [V_{\rm dis}] = \exp [- \int d{\bf x}
V_{\rm dis}^2 ({\bf x})/(2 \kappa) ] / A $, with normalization
constant $ A = \int {\cal D} [V_{\rm dis}] \exp [- \int d{\bf x}
V_{\rm dis}^2 ({\bf x})/(2 \kappa) ], $ which leads to the
configurational average $\langle V_{\rm dis} ({\bf x}) V_{\rm dis} (
{\bf x}^\prime ) \rangle  = \kappa\, \delta ({\bf x} - {\bf
x}^\prime)$.

To derive the effective action for a fixed configuration of
disorder, we define the local chemical potential $\mu ({\bf x}) =
\mu - V_{\rm dis} ({\bf x})$ and follow the functional integral
formulation of the evolution from BCS to BEC
superfluidity~\cite{sademelo-93} to obtain $ S_{\rm eff} = S_0 +
S_{\rm G}, $ where $S_0$ is the action of unbound fermions in the
presence of weak disorder and is given by
\begin{equation}
\label{eqn:effective-action-zero}
S_0 [V_{\rm dis}] =  - \frac{2}{V} \int d{\bf x} \sum_{\bf k}
\ln \left[ 1 + e^{-\xi ({\bf k}, {\bf x})/T} \right],
\end{equation}
with $\xi ({\bf k}, {\bf x}) = \epsilon_{\bf k} - \mu({\bf x})$,
and $\epsilon_{\bf k} = k^2/2m$. The second contribution to
$S_{\rm eff}$ corresponds to Gaussian pairing fluctuations
\begin{equation}
\label{eqn:effective-action-gaussian}
S_{\rm G} [ V_{\rm dis} ] =
\frac{1}{TV}
\int d{\bf x}
\sum_{q} {\bar \Delta} (q) \Gamma^{-1} (q, V_{\rm dis}) \Delta (q),
\end{equation}
where $\Delta (q)$ is the pairing field, % with dimensions of energy,
and
\begin{equation}
\label{eqn:pair-correlation}
\Gamma^{-1} (q, V_{\rm dis})
= \sum_{\bf k}
\frac{ X_1 + X_2 }
{\left[ 2 (iq_{\ell} - \xi_1 - \xi_2  - 2 V_{\rm dis} ({\bf x})) \right] } + C
\end{equation}
is the pair correlation function in the presence of
disorder, where
$C = - (mV/4\pi a_s) + \sum_{\bf k} (2 \epsilon_{\bf k})^{-1}$,
$q = ( {\bf q}, iq_{\ell} )$ represents the four-momentum,
the function
$
X_1 =
\tanh
\left[ \left( \xi_1  + V_{\rm dis} ({\bf x}) \right)/2T \right] $
describes the occupation of fermions
with energy $\xi_1 = \xi ( {\bf k} - {\bf q}/2 )$,
and the function
$
X_2 =  \tanh
\left[ \left( \xi_2  + V_{\rm dis} ({\bf x}) \right)
/2 T \right]
$
describes the occupation of fermions with
energy $\xi_2 = \xi ( {\bf k} + {\bf q}/2 )$.

The description above corresponds to a semiclassical approximation
which is valid for weak disorder potentials.
However, quenched (as opposed to annealed) disorder is
characterized by a static disorder potential and the
necessity to average the thermodynamic potential $\Omega (V_{\rm dis})$ % (or free energy)
rather than the partition function~\cite{belitz-94}.
The configurationally averaged thermodynamic potential is
$\Omega_{\rm av} (\kappa) = \langle \Omega (V_{\rm dis}) \rangle$,
and depends on disorder via the parameter $\kappa$.
%
%%which has dimensions of squared energy times volume.
%
The thermodynamic potential for a fixed configuration is
$\Omega (V_{\rm dis})  = - T \ln Z (V_{\rm dis}) $ when expressed in
terms of the partition function
$Z (V_{\rm dis}) = Z_0 (V_{\rm dis}) \times Z_{G} (V_{\rm dis})$,
where $Z_0 (V_{\rm dis}) =  \exp \left[ - S_0 (V_{\rm dis}) \right]$
is the partition function for unbound fermions, and
$ Z_{G} (V_{\rm dis}) =
\int {\cal D}[\bar \Psi, \Psi]
\exp \left[- S_{G} ( {\bar \Psi}, {\Psi}, V_{\rm dis})
\right]
$
is the partition function for the pairing field.
Thus, $\Omega (V_{\rm dis}) = \Omega_0 (V_{\rm dis})  + \Omega_{G} (V_{\rm dis})$,
where the unbound fermion thermodynamic potential is
$\Omega_0 (V_{\rm dis}) = - T \ln Z_0 (V_{\rm dis}) = T S_0 (V_{\rm dis})$
while the pairing field contribution is
$\Omega_{G} (V_{\rm dis}) = - T \ln Z_{G} (V_{\rm dis})$,
which is approximated by
\begin{equation}
\label{eqn:thermo-gaussian} \Omega_{G} (V_{\rm dis}) = -\frac{T} {V}
\int d{\bf x} \sum_{q} \ln \left[T\, \Gamma (q, V_{\rm dis} ({\bf
x})) \right].
\end{equation}

Expanding in $V_{\rm dis}$ and taking the configurational average leads to
$
\Omega_{0, {\rm av}} (\kappa) = T \langle S_0 [ V_{\rm dis} ] \rangle =
\Omega_{0, {\rm av}} (0) + \Delta \Omega_{0, \rm av} (\kappa)
$
for the thermodynamic potential of unbound fermions,
and
$
\Omega_{G, {\rm av}} (\kappa) =
\Omega_{G, {\rm av} } (0)
+ \Delta\Omega_{G, {\rm av}} (\kappa)
$
for the thermodynamic potential due to the pairing field.
Here, $\Omega_{0, {\rm av}} (0)$ and $\Omega_{G, {\rm av} } (0)$ are just the
thermodynamic potentials without disorder~\cite{sademelo-93},
while the disorder-dependent corrections to the thermodynamic potential are
$
\Delta \Omega_{0, {\rm av}} (\kappa) =
-T \sum_{\bf k} L({\bf k}) \kappa
$,
with
$
L({\bf k}) = 1/\left[ 2 T^2 ( 1 + \cosh (\xi_{\bf k}/T) ) \right]
$,
and
$
\Delta \Omega_{G, {\rm av}} (\kappa) =
- T \sum_{q} M ( q ) \kappa
$
with $M (q) =  P(q) - \vert Q (q) \vert^2/2$.
The first term of $M (q)$ can be expressed as
$
P (q) = \sum_{{\bf k}} \sum_{n = 0}^{3}
c_{n} \left[ X_{1}^n + X_{2}^n \right]\Gamma_0 (q)/V
$
%
%%has dimensions of inverse volume times inverse squared energy
%
where $\Gamma_0 (q) = \Gamma (q, V_{\rm dis} = 0)$
and $ X_{m} = \tanh\left[(\xi_{m})/2T\right] $.
In addition, the coefficients found in the expansion are
$c_{0} = - c_2 = 1/(2 T z)$, $c_{1} = (16T^2 - z^2)/(8T^2 z^3)$,
and $c_3 = 1/(8T^2 z)$, with $z = iq_{\ell} - \xi_1 - \xi_2$.
The second term is
$
Q(q) = \sum_{\bf k} \sum_{n = 0}^{2} d_{n}
\left[ X_{1}^n + X_{2}^n \right] \Gamma_0 (q)/V
$
where the coefficients are $d_0 = -d_2 = 1/(4T z)$,
$d_1 = 1/z^2$.

The total thermodynamic potential is then
$\Omega_{\rm av} (\kappa)  = \Omega_{0, \rm av} (\kappa) +
\Omega_{G,{\rm av}} (\kappa)$, and the corresponding number
equation is $N = -\partial \Omega_{\rm av} (\kappa)/\partial \mu$,
where
\begin{equation}
\label{eqn:number-disorder}
N = N_0 (\kappa) + N_G (\kappa),
\end{equation}
with $N_0 (\kappa) = \partial \Omega_{0,\rm av} (\kappa)/\partial \mu$
being due to unbound fermions,
and $N_G (\kappa) = \partial \Omega_{G,\rm av} (\kappa)/\partial \mu$
being due to pairing.
In the limit of $\kappa \to 0$, the standard results for the thermodynamic
potential, number equation, and time-dependent Ginzburg-Landau
theory are found~\cite{sademelo-93}. However, in the presence of disorder
and in the vicinity of $T_c$, the low-energy and long-wavelength Lagrangian density
${\cal L}_G = {\cal L}_{ND} + {\cal L}_{D}$ derived from the
action of Eqs.~(\ref{eqn:effective-action-zero})
and~(\ref{eqn:effective-action-gaussian}) has two contributions.
The first is a non-dissipative part
$$
{\cal L}_{ND} ({\bf x}, \tau) =  {\bar \Psi} \left[ \partial_{\tau}
- \frac{\nabla^2}{2 m_{*}} - \mu_{*} + \gamma V_{\rm dis} ({\bf x})
\right] \Psi + \frac{g_{*}}{2} \vert \Psi \vert^4,
$$
where $\Psi = \Psi ({\bf x}, \tau)$. Here, the term containing the effective mass
$m_{*}$ is the kinetic energy of the pairing
field, $\mu_{*}$ plays the role of the pairing field chemical potential, $g_{*}$ is
the effective interaction, and $\gamma V_{\rm dis} ({\bf x})$ is the effective
disorder potential.
The second term is the dissipative contribution reflecting the
decay of fermion pairs into unbound fermions, which
takes the Caldeira-Leggett form
$$
{\cal L}_{D} ({\bf x}, \tau) = \frac{\lambda}{2\pi}\int
d\tau^{\prime} \frac {\vert \Psi ({\bf x}, \tau) - \Psi ( {\bf x},
\tau^{\prime}) \vert^2} {(\tau - \tau^{\prime})^2}.
$$

Since we are considering the case of weak disorder,
these parameters can be easily related to original parameters of the
theory without disorder~\cite{sademelo-93}.
Using the notation $X = \tanh(\xi_{\bf k}/2T)$ and
$Y = {\rm sech}^2 (\xi_{\bf k}/2T)$,
the original coefficients are
$
a (\mu, T) = -\frac{m V}{4\pi a_s}
+ \sum_{\bf k} \left[ \frac{1}{ 2 \epsilon_{\bf k} }
-  \frac{X} { 2\xi_{\bf k} } \right],
$
corresponding to the order parameter equation in the absence of disorder when
$a(\mu, T) = 0$,
$$
c (\mu, T) =
\sum_{\bf k}
\left[
\frac{X}{8\xi_{\bf k}^2}
- \frac{Y}{16\xi_{\bf k} T}
+ \frac{XY}{T^2} \frac{k_z^2} {16 m \xi_{\bf k} }
\right],
$$
and $d (\mu, T) =  d_R + i d_I$, with
$d_R = \sum_{\bf k} {X}/{4 \xi_{\bf k}^2} $
and $d_I = \left[ \pi N(\epsilon_F) \sqrt{\mu}/(8 T \sqrt{\epsilon_F} )
\right] \Theta (\mu)$ determine
the effective mass $m_*$, while
$
b (\mu, T) =
\sum_{\bf k}
\left[ \frac{X}{4 \xi_{\bf k}^2 } - \frac{Y^2}{ 8\xi^2_{\bf k} T} \right]
$
and $d$ determine the effective interaction $g_{*}$.
For instance, the effective mass is $m_{*} = d_R m/c$, the effective chemical
potential $\mu_{*} = - a/d_R$, the effective interaction is $g_{*} = b/d_R^2$,
the dimensionless amplitude of the disorder potential is
$\gamma = - (\partial a/\partial \mu)/d_R$, and $\lambda = d_I/d_R$.
A natural choice of dimensionless parameters are
$1/(k_F a_s)$ for interactions,
$\eta = \kappa n_F/\epsilon_F^2$ for disorder, and
${\widetilde T} = T/\epsilon_F$
for temperature, where $k_F$ is the Fermi momentum, $n_F = k_F^3/3\pi^2$ is
the fermion density and $\epsilon_F = k_F^2/2m$ is the Fermi energy.

Notice that the condition
\begin{equation}
\label{eqn:order-parameter-disorder}
\mu_{*} ( {\widetilde T}, 1/(k_F a_s) ) = 0
\end{equation}
corresponds precisely to the order parameter equation in the absence of disorder
$a (\mu, T) = 0$, indicating that this equation is not explicitly affected by
weak disorder, as required by Anderson's theorem~\cite{anderson-59}.
However, the determination of the critical temperature $T_c (\eta)$
and the chemical potential $\mu_c (\eta)$ as a function of
dimensionless disorder $\eta$ for a fixed scattering parameter $1/(k_F a_s)$
requires the simultaneous solutions of the
number Eq.~(\ref{eqn:number-disorder}), and the
order parameter Eq.~(\ref{eqn:order-parameter-disorder}).
\begin{figure} [htb]
\centering
\begin{tabular}{cc}
\epsfig{file = 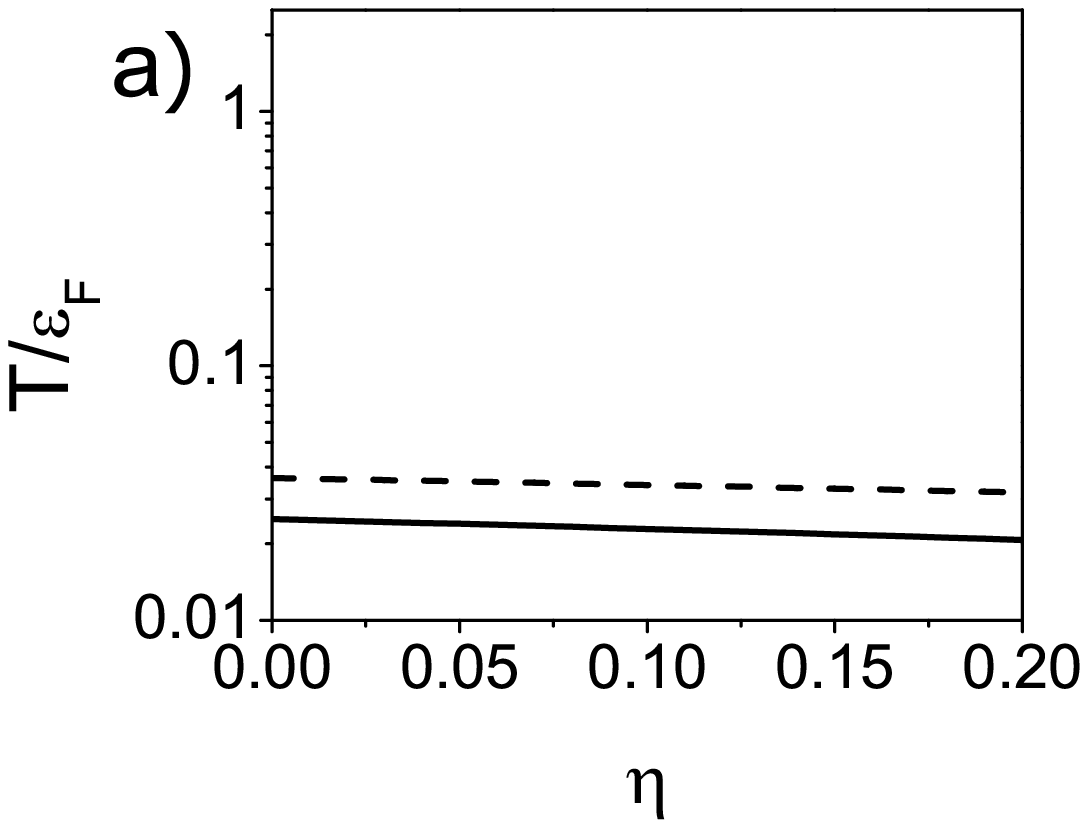, width=0.5\linewidth,clip=} &
\epsfig{file = 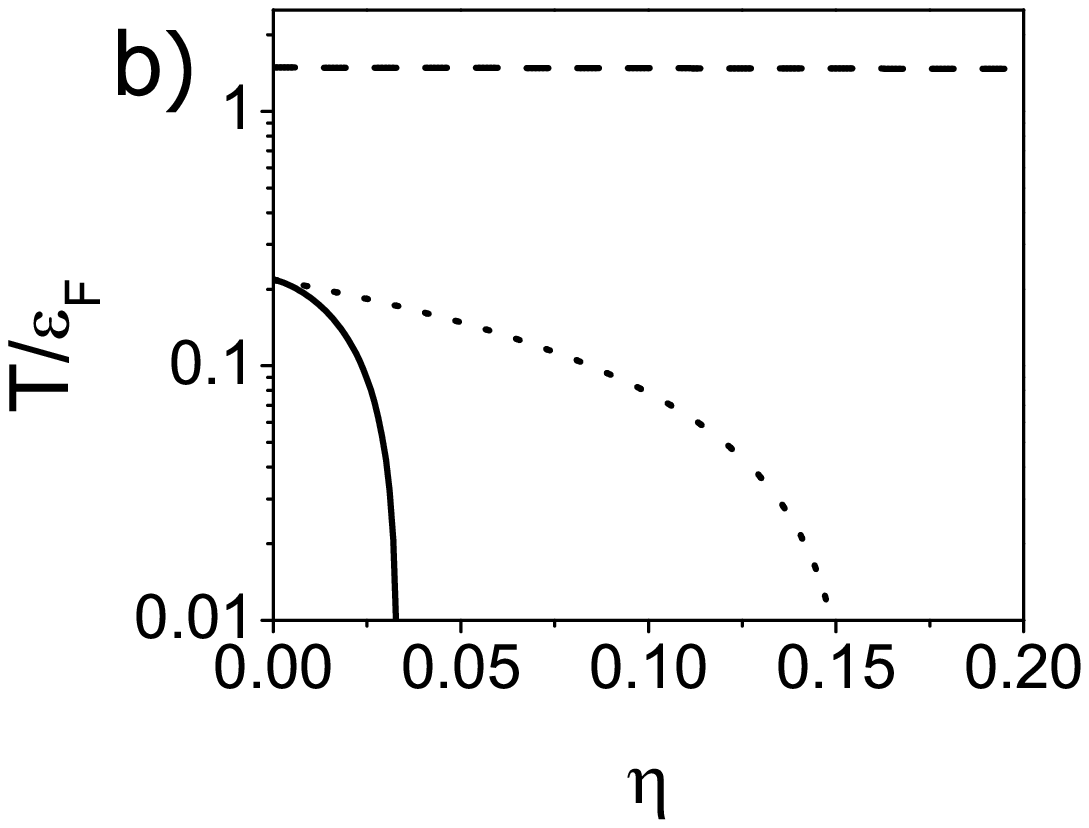, width=0.5\linewidth,clip=}
\end{tabular}
\caption{
\label{fig:tc-disorder}
The pairing temperature $T_p$ (dashed line) and the critical
temperature $T_c$ (solid line) for uniform superfluidity in units of
$\epsilon_F$ as a function of disorder parameter $\eta$
on a) the BCS side for $1/k_F a_s = -1.8$ and on b) the BEC side
for $1/k_F a_s = 1.2$. The dotted line in b) corresponds to the temperature
where the superfluid density vanishes.
}
\end{figure}

In the BCS limit, $\gamma \to 2$ indicating that the pairing field
feels an effective disorder potential twice as large as that felt by
individual unbound fermions, while $\lambda = \pi/(2{\widetilde
T}_c) \gg 1$ indicating that the dynamics of the pairing field is
highly overdamped due to the decay of fermion pairs into unbound
fermions. In the BCS limit without disorder the order parameter
equation alone $\mu_{*} = 0 $ (or $a = 0$) determines the critical
temperature~\cite{sademelo-93}. As seen above this equation remains
unchanged in the presence of weak disorder. Thus, the critical
temperature is essentially unchanged, because the fermion chemical
potential $\mu$ is very large, positive, and remains {\it pinned} to
$\epsilon_F$, in the case of {\it perfect} particle-hole symmetry.
This is a manifestation of Anderson's theorem, and changes in the
critical temperature can occur only via the disorder dependent shift
in the chemical potential, which leads to a change in the single
particle density of states.

The number equation in the BCS limit is approximated by
$N \approx N_0 (0) + \Delta N_0 (\eta)$. The first coefficient
represents the contribution in the absence of disorder
$N_0 (0) = \sum_{\bf k} (1 - X)$, while
the second contribution involves the effects of disorder
$\Delta N_0 (\eta)  =  \eta \sum_{\bf k} X Y/4 (\widetilde T)^2$.
In the case of perfect particle-hole
symmetry $\Delta N_0 (\eta)  =  0$, and indeed
$\mu = \epsilon_F$. Thus, strictly in the
BCS limit $(1/k_F a_s \to -\infty)$,
where the assumption of perfect particle-hole symmetry
is made, the chemical potential is {\it pinned} to the Fermi
energy $(\mu = \epsilon_F)$ and the critical temperature
is unchanged $T_c = (e^{\gamma}/\pi) 8 e^{-2} \exp (-\pi/2k_F|a_s|)$.
Since the presence of disorder does not
alter the total density of fermions
$n_F = k_F^3/3\pi^2$, relaxing the condition of perfect particle-hole symmetry
makes the chemical potential adjust itself to
$\mu = \epsilon_F - \Delta\mu (\eta)$, with
$\Delta \mu (\eta) > 0$
leading to a corresponding reduction
in the critical temperature $T_c (\eta) = T_c (0) - \Delta T_c (\eta)$,
with $\Delta T_c (\eta) > 0$.
In Fig.~\ref{fig:tc-disorder}a, the behavior of
the pairing temperature $T_p (\eta)$ and
the critical temperature $T_c (\eta)$ are shown
in the BCS regime as a function of $\eta$
for $1/k_F a_s = -1.8$.
These curves are obtained by solving the
corresponding number and order parameter equations
for varying $\eta$.

\begin{figure} [htb]
\centerline{\scalebox{0.50}{\includegraphics{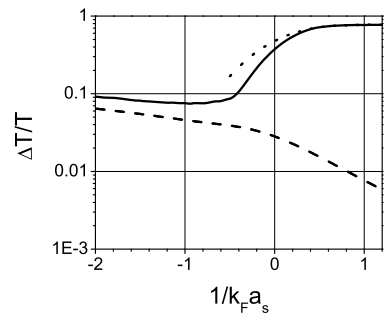} } }
\caption{
\label{fig:tc-relative-change}
The relative change $\Delta T_c /T_c (0)$ (solid line) and
$\Delta T_p/ T_p (0)$ (dashed line)
throughout the BCS to BEC evolution
for $\eta = 0.1$. Notice the minimum in $\Delta T_c/T_c (0)$
occuring in the intermediate region $-1 < 1/k_F a_s < 1$
indicating that the superfluid is more robust to disorder in this
region, while the $\Delta T_p/ T_p(0)$ always decreases indicating that the
pairing temperature becomes less sensitive to disorder in the BEC regime.
The dotted line corresponds to the analytical result in the BEC limit.
}
\end{figure}

As the attraction is increased towards unitarity and $\mu = 0$,
the relative change in $T_c$ for fixed disorder parameter $\eta$,
expressed as $\Delta T_c/T_c (0) = \left[ 1 - T_c (\eta)/T_c (0) \right]$,
decreases as the attraction strength is increased, and reaches a
minimum as indicated in Fig.~\ref{fig:tc-relative-change}.
Beyond this minimum $\Delta T_c/ T_c (0)$ increases again,
being the largest in the BEC regime. These results indicate
that superfluidity is more robust to disorder in
the intermediate regime of $-1 < 1/(k_F a_s ) < 1$,
where the coherence length reaches a minimum~\cite{sademelo-93}
(or the critical current reaches a maximum).
However, the relative change
$\Delta T_p/T_p (0) = \left[ 1 - T_p (\eta)/T_p (0) \right]$
for fixed disorder always decreases with $1/(k_F a_s)$ indicating that
it becomes increasingly more difficult to break pairs with
larger binding energy as the BEC regime is approached.
As discussed next, the difference in behavior between $\Delta T_c/T_c (0)$ and
$\Delta T_p/T_p (0)$ in the BEC regime is attributed
to temporal phase fluctuations, which become increasingly
more important as $1/(k_F a_s)$ increases.

In the BEC regime, $\gamma \to 2$ indicating that the pairing field
feels an effective disorder potential twice as large as that felt by
individual unbound fermions, but $\lambda \to 0$ indicating that the
dynamics of the pairing field is undamped, reflecting that
tightly-bound fermion pairs have very long lifetimes. In the BEC
regime $(1/k_F a_s \gg 1)$ the number equation is dominated by the
contributions coming from $N_{\rm G} \gg N_0$ given that all
fermions are paired into molecular bosons. Here, $N_{G}$ is well
approximated by the number of bound fermions (molecular bosons)
$N_{\rm B}$, since the number of scattering states $N_{\rm SC}$ is
very small. The configurational average can be performed using the
replica technique~\cite{belitz-94,lopatin-02} on the Lagrangian
density ${\cal L}_G$ leading to the action $ S_{G} = (1/T)\sum_{q}
{\bar b} (q) G^{-1}(q) b (q), $ with $ G^{-1}(q) = \frac{q^2}{2m_*}
- \mu_* - i q_{\ell} - \Sigma (q), $
where the self-energy is % of the molecular bosons is
$$
\Sigma (q) = g_*n_B  +
\gamma^2 \kappa  \sum_{\vert {\bf k} \vert < \vert {\bf k}_{c} \vert } \frac { m_*} {k^2 V}
- \gamma^2 \kappa \frac{2m_*^{3/2}}{4\pi}
\sqrt{ \vert \bar\mu_* \vert - iq_{\ell} }.
$$
Here, $G (q)$ plays the role of the molecular boson propagator and leads
to the thermodynamic potential
$
\Omega_{\rm av} \approx
\Omega_{G,{\rm av}} = -  2 T \sum_q \left[ \ln G^{-1} (q)/T \right]
$
with total number of fermions
$N \approx 2 N_{\rm B}$,
where $N_B = T \sum_q G_{\rm av} (q)$,
and renormalized chemical potential
$
\bar \mu_* = \mu_* + g_*n_B  + \gamma^2 \kappa
\sum_{\vert {\bf k} \vert < \vert {\bf k}_{c} \vert }
\left[  m_*/k^2 V  \right]
$.
Upon summation over Matsubara frequencies $iq_{\ell}$, we arrive
to the boson density $n_B = N_B/V$ corrected by disorder
\begin{equation}
\label{eqn:number-bosons}
n_B = \zeta (3/2) (m_{*} T_c/2\pi)^{3/2} + \gamma^2 \kappa T_c m_{*}^3/4\pi^2,
\end{equation}
when $\bar \mu_{*} = 0$. The solution for $T_c$ can be obtained
numerically, but an analytical solution is possible when $\eta =
\kappa n_F/\epsilon_F^2 \ll 1$. This leads to $T_c = T_c (0) \left[
1 - \gamma^2 \kappa T_c (0) m_{*}^3/6\pi^2 n_B \right]$, where $T_c
(0) = 2\pi \left[ n_B/\zeta (3/2) \right]^{2/3}/m_{*}$ is the
Bose-Einstein condensation temperature for a gas of molecular
bosons, which becomes $T_c (0) = 0.218 \epsilon_F$, when $m_{*} \to
2 m$ and $n_B \to n_F/2$.

Unlike in the BCS regime, the critical temperature $T_c$
is directly affected by the presence of disorder in the BEC regime,
and is determined essentially by the number equation.
A small amount of disorder affects the phase coherence of the
molecular bosons through the emergence of an incoherent
part in the molecular boson Green's
function $G^{-1} (q)$
manifested by the branch cut $\sqrt{ \vert \bar\mu_* \vert - iq_{\ell} }$
present in the self-energy $\Sigma (q)$.
If only the zero frequency $(i q_{\ell} = 0)$ contribution
were considered, one would be led to the conclusion
that disorder would not affect the
critical temperature in the BEC regime. However, disorder introduces
quantum (temporal) phase fluctuations in the BEC regime, where
fermions are largely non-degenerate, particle-hole
symmetry is absent, and Anderson's theorem does not apply,
thus leading to a strong suppression of $T_c$.

\begin{figure} [htb]
\centerline{\scalebox{0.50}{\includegraphics{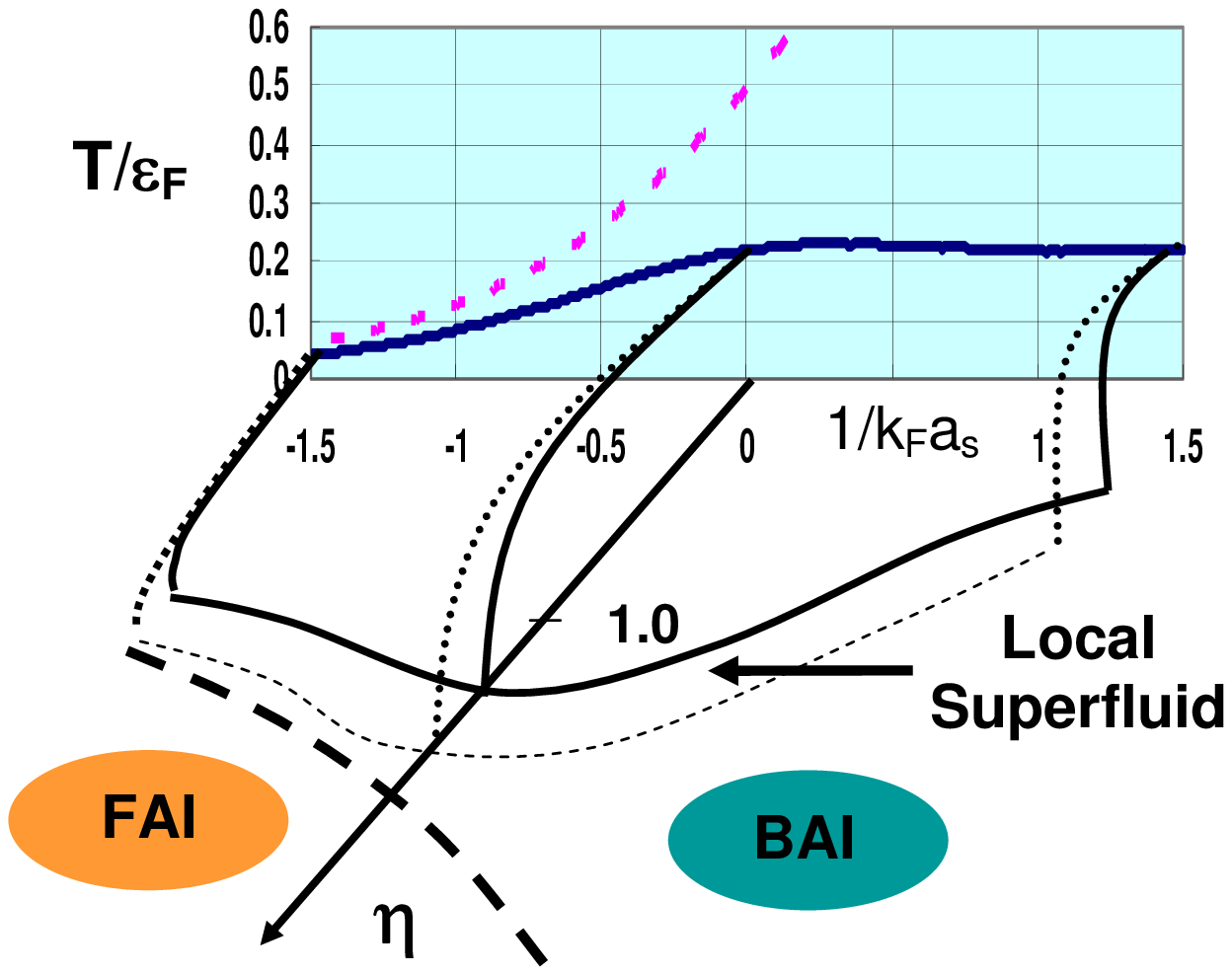} } }
\caption{
\label{fig:phase-diagram}
A schematic finite temperature phase diagram as a function of dimensionless
disorder $\eta$ and interaction $1/k_F a_s$ parameters based on
extrapolations for arbitrary disorder.  Notice existence of regions
of local superfluidity, without global superfluid order, which emerge at intermediate
disorder before the local superfluidity is also destroyed by localization
phenomena. Regions of Fermi-Anderson insulator (FAI) and Bose-Anderson insulator (BAI)
are also indicated.
}
\end{figure}

The critical temperature calculated corresponds
to the transition between the superfluid
and a non-uniform phase containing local superfluid
islands separated from each other by peaks of the disorder potential.
The transition from the local superfluid regime to the normal phase
may be determined by imposing the condition that the
superfluid density $\rho ({\widetilde T}, \eta) = 0$,
which leads to the schematic phase diagram in Fig.~\ref{fig:phase-diagram}.

%%
%

%
%% Conclusions
%
We analyzed the effects of disorder on the critical temperature for
superfluidity of ultracold fermions during the evolution from the
BCS to the BEC regime.
For s-wave superfluids, we showed that weak disorder does not affect the
critical temperature of a BCS superfluid with perfect particle-hole symmetry
in accordance with Anderson's theorem, as the breaking of fermion pairs and the
loss of phase coherence must occur at the same temperature.
However in the BEC regime,
phase coherence is more easily destroyed by disorder without the need of
simultaneously breaking fermion pairs.
Finally, we also showed that the superfluid is more robust to the presence
of weak disorder in the intermediate (crossover) region.
\acknowledgements{We thank NSF (DMR-0709584) for support.}

\end{document}